# Robustness of Network of Networks with Interdependent and Interconnected links

Gaogao Dong[1,2,*], Lixin Tian[1], Ruijin Du[1,3], and H. Eugene Stanley[2]

[1]*Nonlinear Scientific for Research Center, Faculty of Science, Jiangsu University, Zhenjiang, 212013, China*

[2]*Center for Polymer Studies and Department of Physics, Boston University, Boston, Massachusetts 02215 USA*

[3]*College of Mathematics Science, Chongqing Normal University, Chongqing, 401331, China*

**Abstract:**

Robustness of network of networks (NON) has been studied only for dependency coupling (J.X. Gao et. al., Nature Physics, 2012) and only for connectivity coupling (E.A. Leicht and R.M. D'Souza, arxiv:0907.0894). The case of network of n networks with both interdependent and interconnected links is more complicated, and also more closely to real-life coupled network systems. Here we develop a framework to study analytically and numerically the robustness of this system. For the case of starlike network of n Erdös-Rényi (ER) networks, we find that the system undergoes from second order to first order phase transition as coupling strength $q$ increases. We find that increasing intra-connectivity links or inter-connectivity links can increase the robustness of the system, while the interdependency links decrease its robustness. Especially when $q=1$, we find exact analytical solutions of the giant component $P_\infty$ and the first order transition point $p_c^I$. Understanding the robustness of network of networks with interdependent and interconnected links is helpful to design resilient infrastructures.

# I Introduction

In recent years, much progress has been made in the field of complex networks [1-19]. Most of the research have focused on isolated networks that do not connect with or depend on other networks [1-15]. However, most real-world infrastructures are not isolated, are often interconnected, or interdependent, or both. Leicht and D'Souza [20] studied the percolation of interacting networks by introducing a multi-dimension generating function, and found that the interconnected links make the system more robust. Three years ago, robustness of two coupled interdependent networks have been investigated based on percolation theory [21,22]. They found that the system becomes extremely vulnerable because of the dependency coupling. Later, Gao et al. [23,24] developed a generalized framework to study percolation of the "network of networks" (NON). Their findings show that the percolation theory of a single network is a limiting case of a more general case of percolation of interdependent networks. However, in real scenarios, specific nodes (hubs) in one network are not always against random attacks, but malicious attacks. For subsequent study, the above theories on robustness of interdependent networks under initial random attacks have been made significant extention to targeted-attack case very recently [25,26,27]. As we known, in interdependent networks, when nodes in one network fail, the interdependent links carry the failure to nodes in the other networks, and this may happen recursively, and cause the networks splitting into many clusters. Only nodes belonging to the giant cluster of the network are still functional. But for systems in our real life, there they usually contain both two types of inter-links, inter-connectivity links and inter-dependency links. When the cascading failure happen, those small clusters disconnected with the giant component in one network can still be functional through interconnected links, that connecting them to the giant component of other networks. This is one of the reasons why the real-world networks are not that easy to collapse. The percolation of two partially coupled networks with both interdependent and interconnected links has been studied by Hu et al. [28].

*gago999@126.com

In this work, we study analytically and numerically the percolation of a star-like NON under no-feedback condition, which is interdependent and interconnected coupled [see Fig. 1].

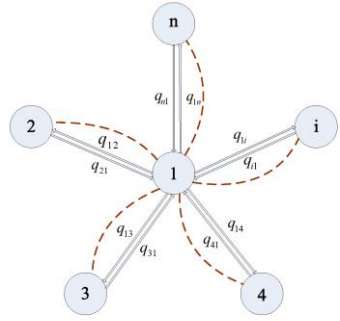

FIG. 1: A star-like NON with interdependent and interconnected links. Each circle represents one network. Full arrows (dependency links) pointing from nodes in network $1$ to nodes in network $i$, indicate that a certain fraction $q_{1i} > 0$ of nodes of network $i$ directly depend on nodes of network $1$. Dashed line (connectivity links) connect network $1$ and $i$ randomly.

**II General formalism**

Our NON system are coupled by both dependency and connectivity links. Each node of the NON represent a network. We suppose that each network $i$ $(i=1,2,...,n)$ consists of $N_i$ nodes linked together by connectivity links. Two networks $i$ and $j_l$ are partially coupled by dependency links, so that a fraction $q_{j_l i}$ of nodes in network $i$ depends on nodes in network $j_l$, and a fraction $q_{ij_l}$ of nodes in $j_l$ depends on nodes in network $i$, where $l=1,2,...,m_i, i=1,2,...,n$. When $q_{ij_l} = q_{j_l i} = 1$, the partially dependent pair becomes fully dependent. We further assume that each node $a$ in network $i$ depends on only one node $b$ in network $j_l$ (uniqueness condition), and if node $a$ in network $i$ depends on node $b$ in network $j_l$, and node $b$ in network $j_l$ depends on node $c$ in network $i$, node $a$ must coincide with node $c$ (no-feedback condition) [29]. In addition, the connectivity links within network $i$ and between networks $i$ and $j_l$ can be described by degree distribution $\rho_{k_{i1},......,k_i,...,k_{ij_l},...,k_{in}}$, which denotes the probability of an $i$-node to have $k_i$ links to other $i$-nodes and $k_{ij_l}$ links towards $j_l$-nodes, where $l=1,2,...,m_i$. As in Ref. [20], we get a $n$ dimensional

*gago999@126.com

generating function describing all the connectivity links, $\mathcal{G}_0^i(x_1, x_2, \ldots, x_n) =$
$\sum_{\substack{k_{i1},\ldots,k_i,\ldots,k_{ij_l},\ldots,k_{in}, \\ l=1,\ldots,m_i}} \rho_{k_{i1},\ldots,k_i,\ldots,k_{ij_l},\ldots,k_{in}} x_1^{k_{i1}} \cdots x_i^{k_i} \cdots x_{j_l}^{k_{ij_l}} \cdots x_n^{k_{in}}$, where $k_{ih} = 0$, if $ih \neq ij_l, i$. Analogously

we introduce the generating function of the underlying branching processes, $\mathcal{G}_1^{ij_l}(x_1, x_2, \ldots, x_n) =$

$$\frac{\frac{\partial \mathcal{G}_0^i(x_1, x_2, \ldots, x_n)}{\partial x_{j_l}}}{\frac{\partial \mathcal{G}_0^i(1,1,\ldots,1)}{\partial x_{j_l}}}, l = 1, 2, \ldots, m_i, i = 1, 2, \ldots, n.$$

Initially, $1 - p_i$ fraction of nodes are removed from each network $i$. We define $x_i$ as the remaining nodes that survived in network $i$ after removing all nodes affected by the initial failure and those nodes depending on the failed nodes in the other networks. The fraction of nodes in the giant component of each network $i$ is $P_{\infty,i} = x_i g_i(x_1, x_2, \ldots, x_n)$, where $g_i(x_1, x_2, \ldots, x_n)$ is the fraction of the remaining nodes belonging to the giant component of network $i$. The function $g_i(x_1, x_2, \ldots, x_n)$ can be expressed as [30-32]

$$g_i(x_1, x_2, \ldots, x_n) = 1 - \mathcal{G}_0^i(1 - x_1(1 - f_{1i}), \ldots, 1 - x_i(1 - f_i), \ldots, 1 - x_{j_l}(1 - f_{j_l i}), \ldots, 1 - x_n(1 - f_{ni})), \quad (1)$$

where

$$\begin{aligned} f_i &= \mathcal{G}_1^i(1 - x_1(1 - f_{1i}), \ldots, 1 - x_i(1 - f_i), \ldots, 1 - x_{j_l}(1 - f_{j_l i}), \ldots, 1 - x_n(1 - f_{ni})), \\ f_{ij_l} &= \mathcal{G}_1^{ij_l}(1 - x_1(1 - f_{1i}), \ldots, 1 - x_i(1 - f_i), \ldots, 1 - x_{j_l}(1 - f_{j_l i}), \ldots, 1 - x_n(1 - f_{ni})), \end{aligned} \quad (2)$$

and if $hi \neq j_l i, i$, $f_{hi} = 1$.

The cascading process can be described by the following system of $n$ equations:

$$x_i = p_i \prod_{l=1}^{m_i} (1 - q_{j_l i} + q_{j_l i} y_{j_l i} g_{j_l}), \quad (3)$$

where the product is taken over the $m_i$ networks interlinked with network $i$ by partial dependency links, and

$$y_{j_l i} = \frac{x_{j_l}}{1 - q_{ij_l} + q_{ij_l} y_{ij_l} g_i}, \quad (4)$$

has the meaning of the fraction of nodes in network $j_l$ survived after the damage from all the networks connected to network $j_l$ except from network $i$ is taken into account. The damage from network $i$ must be excluded due to the no-feedback condition. In the absence of the no-feedback condition, Eq. (3) becomes much simpler because of $y_{j_l i} = x_{j_l}$.

In this study, we suppose all of the degree distributions of inter and intra networks $i$ and $j_l$ are Poisson distribution [33-35], all of the functions can be more simple. Assume $\bar{k}_i, \bar{k}_{j_l}$ are the average intra-links degrees in networks $i$ and $j_l$, and $\bar{k}_{ij_l}, \bar{k}_{j_l i}$ are the average inter-links

*gago999@126.com

degrees between $i$ and $j_l$. Then,

$$G_0^i(x_i) = e^{\bar{k}_i(x_i-1)}, \quad G_0^{ij_l}(x_{j_l}) = e^{\bar{k}_{ij_l}(x_{j_l}-1)}, l=1,2,\ldots,m_i, \tag{5}$$

and

$$\mathcal{G}^i(x_1,x_2,\ldots,x_n) = \mathcal{G}^{ij_l}(x_1,x_2,\ldots,x_n) = \mathcal{G}_0^i(x_1,x_2,\ldots,x_n) = G_0^i(x_i)\prod_{l=1}^{m_i}G_0^{ij_l}(x_{j_l}). \tag{6}$$

Substituting Eqs. (5) and (6) into Eqs. (1) and (2), we obtain

$$g_i = 1 - e^{-\bar{k}_i x_i g_i - \sum_{l=1}^{m_i}\bar{k}_{ij_l}x_{j_l}g_{j_l}}, \tag{7}$$

$$f_{ij_l} = f_i = 1 - g_i = e^{-\bar{k}_i x_i g_i}\prod_{l=1}^{m_i}e^{-\bar{k}_{ij_l}x_{j_l}g_{j_l}} = e^{-\bar{k}_i x_i g_i - \sum_{l=1}^{m_i}\bar{k}_{ij_l}x_{j_l}g_{j_l}}. \tag{8}$$

**III Star-like NON under no-feedback condition**

For the case of a partially coupled star-like NON [Fig. 1] under no-feedback condition, we have $y_{j1} = p_j (j=2,3,\ldots,n)$, where network 1 is the central network, and network $j$ represents the surrounding network. Under the following simplifying conditions that $p_1 = p_j = p$, $q_{1j} = q_{j1} = q, \bar{k}_1 = \bar{k}_j = \bar{k}, \bar{k}_{1j} = \bar{k}_{j1} = \bar{K}, j=2,3,\ldots,n,$ we have $x_2 = x_3 = \ldots = x_n$, so Eqs. (3) and (8) become

$$\begin{aligned} x_1 &= p(1-qf_2)^{n-1}, \\ x_2 &= 1-q+pq(1-f_1)(1-qf_2)^{n-2}, \end{aligned} \tag{9}$$

and

$$\begin{aligned} f_1 &= e^{-\bar{k}p(1-f_1)(1-qf_2)^{n-1}-(n-1)\bar{K}(1-f_2)\{1-q+pq(1-f_1)(1-qf_2)^{n-2}\}}, \\ f_2 &= e^{-\bar{k}(1-f_2)\{1-q+pq(1-f_1)(1-qf_2)^{n-2}\}-\bar{K}p(1-f_1)(1-qf_2)^{n-1}}. \end{aligned} \tag{10}$$

From the definitions of $P_{\infty,i}$, we get

$$\begin{aligned} P_{\infty,1} &= p(1-f_1)(1-qf_2)^{n-1}, \\ P_{\infty,2} &= (1-f_2)\{1-q+pq(1-f_1)(1-qf_2)^{n-2}\}. \end{aligned} \tag{11}$$

We verify our theory, Eq. (11), by comparing theoretical predictions with simulation results for different coupling strength $q$, as shown in Fig. 2. Additionally, from Fig. 2, we observe that the giant components of the central and surrounding networks undergo from second order to first order phase transition as $q$ increases. In further, by analyzing the graphical solution of Eq. (10), a critical line can be found for different $n$, $\bar{k}$ and $\bar{K}$. When the coupling is strong ($q=0.8$ in Fig. 3), the starlike NON shows a first order phase transition. When the coupling is weak ($q=0.1$ in Fig. 3), the starlike NON exhibits a second order phase transition. Note that for the same $q$, the central network becomes less robust when $n$ increases, but more robust when $\bar{k}$ or $\bar{K}$ increases.

*gago999@126.com

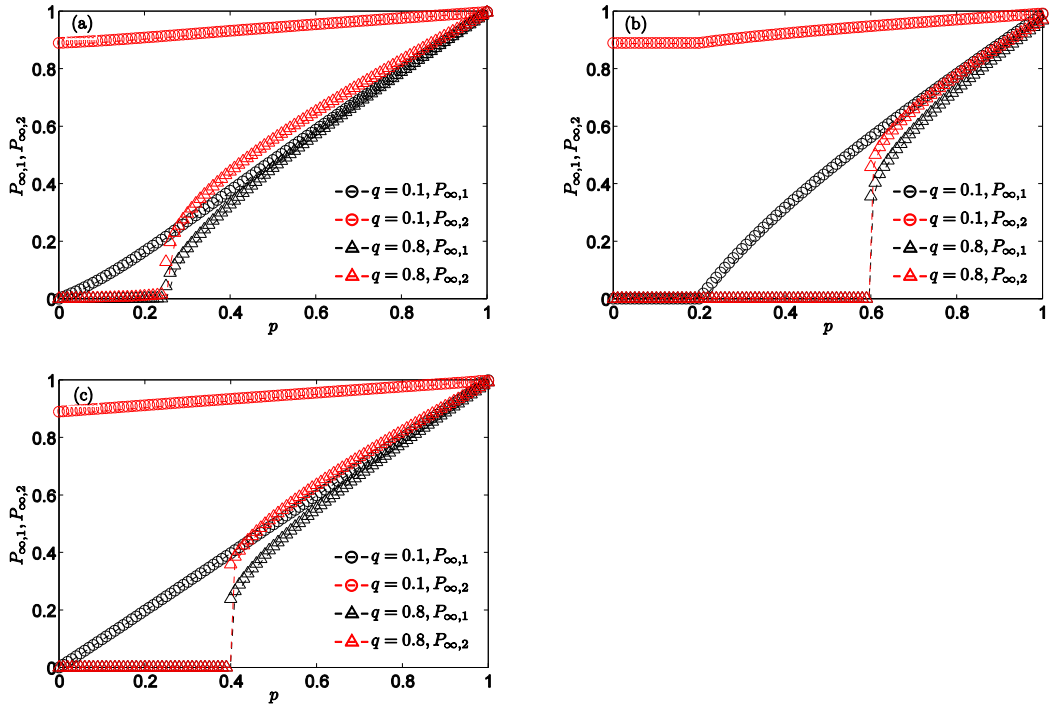

FIG 2. For a starlike network of $n$ $ER$ networks, comparison between simulations (circle and triangle) and theory (dashed line) of $P_{\infty,1}, P_{\infty,2}$ as a function of the initial attack on the central network, $1-p$, for different $q$. (a) $n=2, \bar{k}=5, \bar{K}=1$, (b) $n=5, \bar{k}=5, \bar{K}=0$, (c) $n=5, \bar{k}=5, \bar{K}=1$. In simulation, $N_i=N=10^5$ and the results are averaged over 50 realizations.

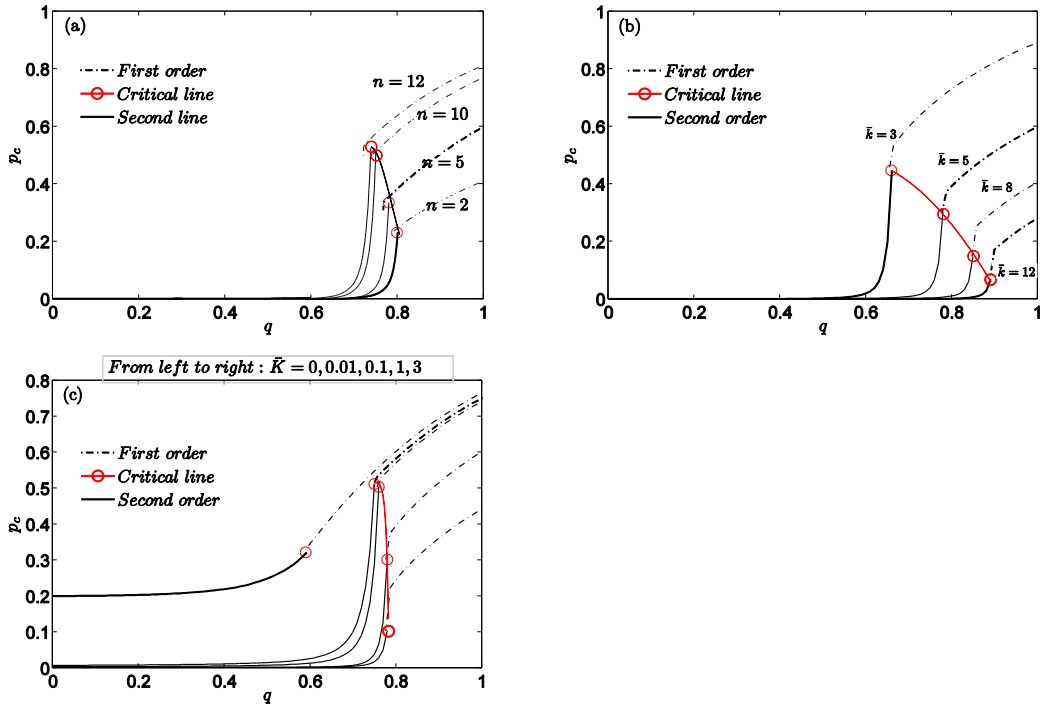

FIG 3. The relation between $p_c$ and $q$. (a) $\bar{k}=5, \bar{K}=1, n=2,5,10,12$, (b) $n=5, \bar{K}=1$, $\bar{k}=3,5,8,12$, (c) $n=5, \bar{k}=5, \bar{K}=0, 0.01, 0.1, 1, 3$. The curves connecting the circles show the critical

*gago999@126.com

lines, above which the system shows a first order phase transition and below which a second order phase transition.

Especially when $q=1$, the starlike NON are fully interdependent. Eqs. (9)-(11) yield simple forms

$$x_1 = p(1-f_2)^{n-1},$$
$$x_2 = p(1-f_1)(1-f_2)^{n-2}, \qquad (12)$$

$$f_1 = e^{-(\bar{k}+(n-1)\bar{K})p(1-f_1)(1-f_2)^{n-1}},$$
$$f_2 = e^{-(\bar{k}+\bar{K})p(1-f_1)(1-f_2)^{n-1}}, \qquad (13)$$

$$P_\infty = P_{\infty,1} = P_{\infty,2} = p(1-e^{-(\bar{k}+(n-1)\bar{K})P_\infty})(1-e^{-(\bar{k}+\bar{K})P_\infty})^{n-1}. \qquad (14)$$

Fig. 4 show excellent agreement between simulations of the giant component and the results in Eq. (14). Furthermore, each network has the same giant component, and shows a first order percolation transition ($n>1$). From Fig. 4(b), we see that $p_c$ increases as $n$ increases. From Fig. 4(a) and 4(c), we obtain that $p_c$ increases as $\bar{k}$ or $\bar{K}$ decreases.

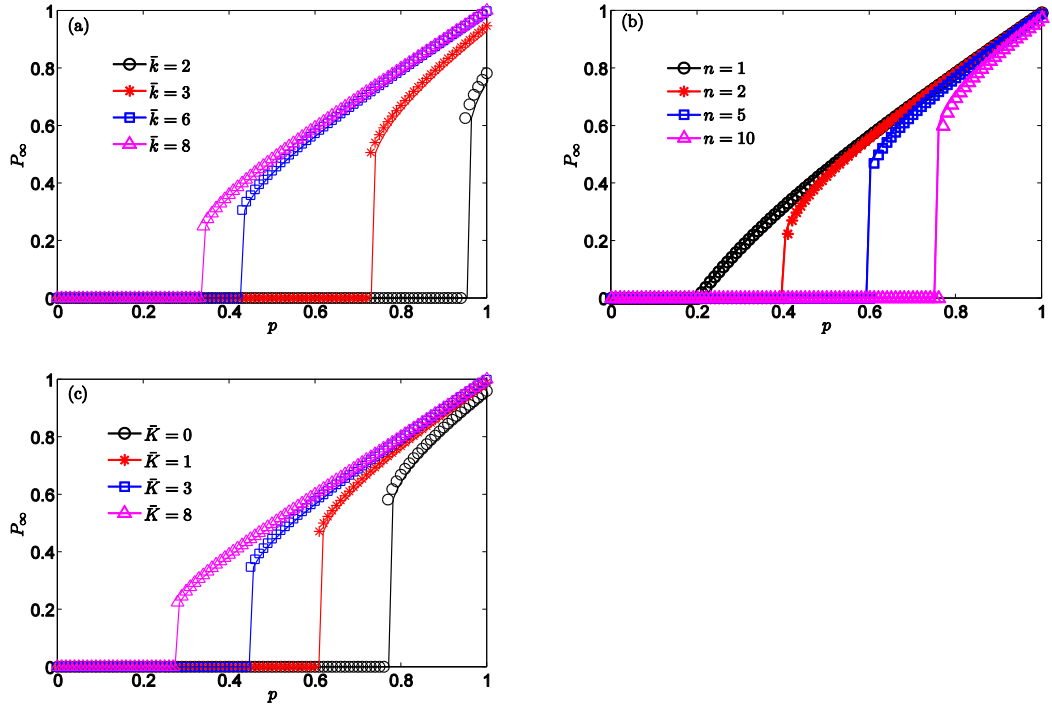

FIG 4. When $q=1$, comparison between simulation (circle, star, square and triangle) and theory (solid line) for the giant component $P_\infty$ as a function of $p$. (a) $n=3, \bar{K}=1, \bar{k}=2,3,6,8$, (b) $\bar{k}=5, \bar{K}=1, n=1,2,5,10$, (c) $n=5, \bar{k}=5, \bar{K}=0,1,3,8$.

Solving the second equation of system (13), we get

$$f_1 = 1 + \frac{\ln f_2}{p(\bar{k}+\bar{K})(1-f_2)^{n-1}} = H_1(f_2). \qquad (15)$$

Substituting Eq. (15) into the first equation of system (13), we obtain

*gago999@126.com

$$f_1 = f_2^{\frac{\bar{k}+(n-1)\bar{K}}{\bar{k}+\bar{K}}} = H_2(f_2). \tag{16}$$

The curves of Eqs. (15) and (16) are tangentially touching at the critical fraction $p = p_c^I$ for the first order phase transition,

$$(\frac{dH_1}{df_2} = \frac{dH_2}{df_2})\big|_{p=p_c^I}. \tag{17}$$

From Eqs. (15)-(17), we get the threshold

$$p_c^I = \frac{1-f_2+(n-1)f_2 \ln f_2}{[\bar{k}+(n-1)\bar{K}](1-f_2)^n f_2^{\frac{\bar{k}+(n-1)\bar{K}}{\bar{k}+\bar{K}}}}, \tag{18}$$

where, $f_2$ satisfies the equation,

$$f_2^{-\frac{\bar{k}+(n-1)\bar{K}}{\bar{k}+\bar{K}}} + \frac{[\bar{k}+(n-1)\bar{K}](1-f_2)\ln f_2}{(\bar{k}+\bar{K})[1-f_2+(n-1)f_2\ln f_2]} = 1. \tag{19}$$

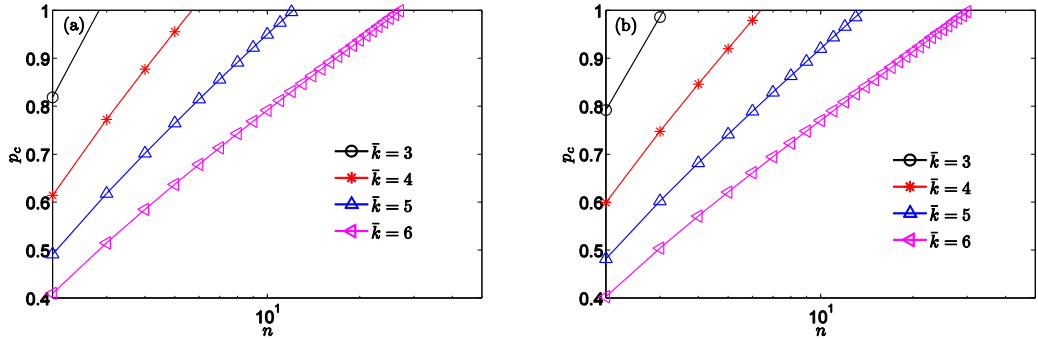

FIG. 5 The critical threshold $p_c$ as a function of $n$ for different $\bar{k}$. (a) $\bar{K}=0$, (b) $\bar{K}=0.1$.

From Fig.5, we can see that $p_c$ increases as $n$ increases or $\bar{k}$ decreases. Comparing Fig.5(a) and 5(b), for the same $n$ and $\bar{k}$, we find that $p_c$ decreases with $\bar{K}$ increasing.

**IV Conclusion**

In summary, we have introduced a formalism to study the robustness of network of networks with interdependent and interconnected links. For $q=0$, our system becomes NON with only interconnectivity links, which has been studied in Ref. [20]. For $\bar{K}=0$, our system becomes NON with only interdependent links, which has been studied in Refs. [23, 24]. For a starlike network of n ER networks, the system exhibits two phase transitions as $q$ changes. There exists a critical line, above which the system shows a first order phase transition, and below which a second order phase transition. For the same $q$, the central network becomes less robust when $n$ increases, but more robust when $\bar{k}$ or $\bar{K}$ increases. In particular, for $q=1$, we get the analytical expression of the first order phase transition point. Our analytical theory is developed for ER networks, but the same qualitative conclusions hold for any network systems topology. In addition, we also consider our framework with feedback condition and has made some progress.

*gago999@126.com

**ACKNOWLEDGMENTS**

L.T. thanks the National Natural Science Foundation of China (Grant Nos. 11171135, 71073071, 71073072, 51276081) and the Major Program of the National Social Science Foundation of China (Grant No. 12&ZD062) for support. M.F. thanks the National Youth Natural Science Foundation of China (Grant No. 71303095) for support, G. D. thanks the Natural Science Foundation of Jiangsu Province (Grant Nos. BK20130535, SBK201342872) and National Natural Science Foundation of China (Grant No. 51305168). HES thanks ONR (Grant Nos. N00014-09-1-0380, N00014-12-1-0548), DTRA (Grant Nos. HDTRA-1-10-1-0014, HDTRA-1-09- 1-0035) and NSF (Grant No. CMMI 1125290) for support.

*gago999@126.com

*gago999@126.com